# DETERMINING THE $H^+$ REGION / PDR EQUATION OF STATE IN STAR FORMING REGIONS

N. P. Abel & G. J. Ferland
University of Kentucky, Department of Physics & Astronomy, Lexington, KY 40506; npabel2@uky.edu, gary@pa.uky.edu

# Abstract

The emission line regions of starburst galaxies and active nuclei reveal a wealth of spectroscopic information. A unified picture of the relationship between ionized, atomic, and molecular gas makes it possible to better understand these observations. We performed a series of calculations designed to determine the equation of state, the relationship between density, temperature, and pressure, through emission-line diagnostic ratios that form in the $H^+$ region and PDR. We consider a wide range of physical conditions in the H+ region. We connect the $H^+$ region to the PDR by considering two constant pressure cases, one with no magnetic field and one where the magnetic field overwhelms the thermal pressure. We show that diagnostic ratios can yield the equation of state for single $H^+$ regions adjacent to single PDRs, with the results being more ambiguous when considering observations of entire galaxies. As a test, we apply our calculations to the Orion $H^+$/PDR region behind the Trapezium. We find the ratio of thermal to magnetic pressure in the PDR to be ~1.2. If magnetic and turbulent energy are in equipartition, our results mean the magnetic field is not the cause of the unexplained broadening in M 42, but may significantly affect line broadening in the PDR. Since Orion is often used to understand physical processes in extragalactic environments, our calculations suggest magnetic pressure should be considered in modeling such regions.

# 1 Introduction

Classically, H II regions, which we refer to as $H^+$ regions, and PDRs, have been treated as distinct problems. In actuality, the two regions are dynamically linked by a continuous flow. This brings up the question of the equation of state, the relationship between density and temperature, since the flow extends from cold molecular gas into hot ionized regions. Magnetic fields also play a role in the equation of state, as the field is coupled to the gas through electromagnetic forces and collisions. Magnetic pressure also balances gravitational pressure, decreasing the rate of star formation (Crutcher, Heiles, & Troland 2003). There are two simple limiting cases; constant gas pressure, where the cool (~100 K)

PDR is ~$10^2$ times denser than the warm (~$10^4$ K) $H^+$ region, and a magnetically dominated geometry, where the densities may be the same.

Many pressure terms affect conditions in the ISM (Henney 2006). Thermal gas pressure is often assumed to dominate in the $H^+$ region. Turbulent and magnetic pressure, often in equipartition, is often dominant in molecular gas. Two approaches can be taken to simulate an $H^+$ region and PDR. In a true MHD calculation the microphysics must be simplified. Codes that do not compromise the microphysics cannot do a full MHD simulation. This study falls into the latter category. Doing both is beyond the capability of today's computers. Here we compute the full spectrum of an $H^+$ region and PDR by treating the two as a single self-consistent problem. This is done in the limit where only thermal and magnetic pressures contribute. While this is a simplification, this does approximate environments where these two pressure terms dominate the equation of state. Examples may include the Orion $H^+$ region and Starbursts galaxies.

This paper seeks to determine whether infrared emission lines can determine the equation of state linking the $H^+$ region to the PDR, and the role of the magnetic field in the equation of state. Here we assume that the cloud is static and that the total pressure is a combination of thermal and magnetic terms, and that the field and gas density is related by a power law. In Section 2, we present the equations governing an $H^+$ region and PDR in overall pressure equilibrium and identify infrared spectral diagnostics that can determine the role of magnetic fields in such an environment. In Sections 3 & 4, we present theoretical calculations of these spectral diagnostics. In Section 5 we show an application of our work to the Orion environment.

# 2 Equation of State in Star-Forming Regions

## 2.1 An $H^+$ Region and PDR in Pressure Equilibrium

If magnetic fields are ignored, ionized and molecular gases are often assumed to be in a state of gas pressure equilibrium (isobaric). This includes many starburst galaxies (Carral et al, 1994; Lord et al 1996) and normal galaxies whose luminosities are dominated by star-formation (Malhotra et al, 2001). In addition, a weak-D ionization front (Henney et al. 2005) is also nearly in constant gas pressure.

Different physical processes dominate the total pressure in an $H^+$ region and PDR. In the $H^+$ region, the dominant pressure is thought to be thermal pressure, $P^{th}=nkT$, owing to the high temperatures (~$10^4$ K) in $H^+$ regions (Ferland 2001). In colder, more molecular environments such as PDRs, magnetic ($P^{mag} = B^2/8\pi$) or turbulent pressure is thought to dominate over $P^{th}$ (Crutcher 1999; Heiles & Crutcher 2005; Tielens & Hollenbach 1985). Magnetic and turbulent pressures are often thought to be in rough equipartition, with ~0.8 considered representative for the ratio $P^{\mathrm{mag}}$ to $P^{\mathrm{turb}}$ (Heiles & Crutcher 2005).

We now consider an $H^+$ region in total pressure equilibrium with a PDR. The equation of pressure equilibrium can be written as:

$$\sum_i P^i_{\mathrm{H(+)}} = \sum_i P^i_{\mathrm{PDR}} \tag{1}$$

If we only consider the effects of $P^{\mathrm{mag}}$ and $P^{\mathrm{th}}$, we get:

$$2P^{\mathrm{mag}}_{(\mathrm{H}^+)} + P^{\mathrm{th}}_{(\mathrm{H}^+)} = P^{\mathrm{th}}_{\mathrm{PDR}} + 2P^{\mathrm{mag}}_{\mathrm{PDR}} \tag{2}$$

In equation 2, we have doubled the magnetic pressure under the assumption that the turbulent and magnetic pressure are in equipartition. Equation 2 neglects the effect of stellar radiation pressure and ram pressure due to microturbulence or bulk motions on the equation of state. Our calculations presented in section 3 include radiation pressure but it is never important. The calculations also assume a static geometry, thereby neglecting ram pressure. The gas is coupled to the magnetic field $B$ though the Lorentz force and collisions. Therefore, $B$ is proportional to some power of density, $B \propto n_{\mathrm{H}}^{\kappa}$ (Henney et al. 2005). Observations (Crutcher 1999) and theoretical calculations (Fiedler & Mouschovias 1993) both suggest that $\kappa = 1/2$, where $n_{\mathrm{H}}$ is the hydrogen density. For the collapse of a spherical cloud with a magnetic field, magnetic flux conservation implies $\kappa = 2/3$ (see Crutcher 1999).

Using $B \propto n_{\mathrm{H}}^{\kappa}$, we find that $B$ in the $H^+$ region and PDR are related by:

$$B(\mathrm{H}^+) = B_{\mathrm{PDR}} \left( \frac{n(\mathrm{H}^+)}{n_{\mathrm{PDR}}} \right)^{\kappa} \tag{3}$$

where $n(\mathrm{H}^+)$, $B(\mathrm{H}^+)$, $n_{\mathrm{PDR}}$, and $B_{\mathrm{PDR}}$ are the hydrogen density and magnetic field strength in the $H^+$ region and PDR.

We are now in a position to derive an equation linking $B$, $n$, and $T$ in the $H^+$ region and PDR. From the definitions of $P^{\mathrm{th}}$ and $P^{\mathrm{mag}}$, along with equations 2 & 3, we have:

$$\left(\frac{B_{\rm PDR}^2}{4\pi}\right)\left(\frac{n({\rm H}^+)}{n_{\rm PDR}}\right)^{2\kappa}+n({\rm H}^+)kT({\rm H}^+)=\frac{B_{\rm PDR}^2}{4\pi}+n_{\rm PDR}kT_{\rm PDR} \quad (4)$$

where $T(\mathrm{H}^+)$ and $T_{\rm PDR}$ is the temperature in the $\mathrm{H}^+$ region and PDR. It is often convenient to define $\frac{P_{PDR}^{th}}{P_{PDR}^{mag}}$ as $\beta$ (see, for instance, Heiles & Troland 2005).

Inserting this definition into equation 4 and solving for $\beta$, we get:

$$\begin{aligned}\beta &= \frac{P_{PDR}^{th}}{P_{PDR}^{mag}} \\ &= 2\frac{1-\left(\frac{n({\rm H}^+)}{n_{\rm PDR}}\right)^{2\kappa}}{\left(\frac{n({\rm H}^+)T({\rm H}^+)}{n_{\rm PDR}T_{\rm PDR}}\right)-1}\end{aligned} \quad (5)$$

where the factor of 2 comes from assuming equipartition between magnetic and turbulent pressure.

Equation 5 yields estimates of $\beta$ and $n_{\rm PDR}$ in environments where gas or magnetic/turbulent pressures are the two contributors to the total pressure. When the magnetic field dominates over the gas pressure in both regions, $n(\mathrm{H}^+) = n_{\rm PDR}$ and $\beta = 0$. When the field is small $P_{\rm H(+)}^{\rm th} = P_{\rm PDR}^{\rm th}$ and $\beta = \infty$. In practice $\beta$ can be between these extremes. Crutcher (1999) found $\beta = 0.04$ for cold molecular regions of the ISM. Typically the temperatures in the two regions are related by $T(\mathrm{H}^+) \approx 10^2 \times T_{\rm PDR}$. If this value of $\beta$ is typical then $n_{\rm PDR} \approx 3 \times n(\mathrm{H}^+)$ rather than the 100:1 ratio found in the constant thermal pressure case.

## 2.2 Determining $\beta$ and $B$ through Infrared Spectroscopy

In this Section, we show how infrared observations can determine the density and temperature in each region, which then determines $\beta$ and $B$. We identify the combination of infrared $\mathrm{H}^+$ region and PDR emission-line diagnostics that are needed to determine $\beta$ and $B$.

### 2.2.1 $n(\mathrm{H}^+)$

Observations of emission lines formed in the ground term of the same ion can determine the electron density $n_e$ (Rubin et al. 1994; Malhotra et al, 2001; Abel et al. 2005, henceforth referred to as A05), which is, to within 10%, equal to the hydrogen density. In the infrared, the average electron energy is much greater than the excitation potential, eliminating the dependence of temperature on the line ratio (Osterbrock & Ferland 2006). However, each line has a different critical density, making the line ratio dependent on $n_e$. Examples of density diagnostics in the infrared are [O III] 51.8 μm / 88.3 μm, [S III] 18.7 μm /33.5 μm, and [N II] 121.7 /205.4 μm.

### 2.2.2 *T*($H^+$), $T_*$, and *U*

As pointed out by Rubin et al. (1994), no direct method exists to determine $T(H^+)$ (which, for $H^+$ regions, is more commonly referred to as $T_e$) using IR diagnostics alone. Dinerstein, Lester, and Werner (1985) showed that the ratio of [O III] infrared to optical emission lines can determine $T(H^+)$. For a wide range of $H^+$ regions with typical abundances, $T(H^+) \approx (0.8\text{-}1)\times10^4$ K.

For a given chemical composition, knowledge of stellar temperature ($T_*$), ionization parameter (*U*, which is the dimensionless ratio of hydrogen ionizing flux $\phi$ to hydrogen density), and $n(H^+)$ we can also predict $T(H^+)$ from theoretical calculations (Shields & Kennicutt 1995). Many elements are observed in multiple ionization stages in the IR. The ratio of their emission-line intensities is sensitive to the shape and intensity of the radiation field, which is set by $T_*$ and *U*. Common $T_*$ and *U* emission-line diagnostic ratios include [Ne III] 15.6μm to [Ne II] 12.8μm, [S IV] 10.8μm to [S III] 18.7μm and [N III] 57.1μm to [N II] 121.7μm. As pointed out by many authors (e.g. Morisset 2004, Giveon et al. 2002) a minimum of two of these emission line ratios is required to independently determine $T_*$ and *U* , for a given stellar atmosphere.

### 2.2.3 $n_{PDR}$ & $T_{PDR}$

Fine-structure line emission from elements with ionization potentials < 13.6 eV, combined with theoretical calculations, can determine $n_{\rm PDR}$ and $T_{\rm PDR}$ (e.g. Wolfire, Tielens, & Hollenbach 1990; Kaufman et al. 1999). Common lines used for this analysis include [C II] 157.6 μm, [C I] 369.7, 609 μm, [O I] 63.2, 145.5 μm, and [Si II] 34.8 μm. Such an analysis can determine $n_{\rm PDR}$ and the intensity of the UV radiation field relative to the interstellar radiation field, parameterized by $G_0$ ($1G_0 = 1.6\times10^{-3}$ ergs cm$^{-2}$ s$^{-1}$; Habing 1968). Knowing $n_{\rm PDR}$ and $G_0$, theoretical calculations then determine $T_{\rm PDR}$. This approach is similar to determining $T(H^+)$ from $n(H^+)$ and the properties of the hydrogen-ionizing continuum.

PDR calculations assume that the $H^+$ region does not contribute to the total fine-structure emission, an assumption which is not always true. In low-density $H^+$ regions, or regions where the size of the PDR is small compared to the $H^+$ region, a significant portion of [C II], [O I], or [Si II] emission can come from the $H^+$ region (Carral et al. 1994; Heiles 1994; A05). The $H^+$ region component to the fine-structure emission must be estimated in order to use PDR calculations to derive $n_{\rm PDR}$ and $T_{\rm PDR}$. Recently, A05 calculated the $H^+$ region contribution [C II], [O I], and [Si II] emission for a wide range of stellar temperatures ($T_*$), ionization parameters (*U*), and $n(H^+)$ using the spectral synthesis code Cloudy (Ferland et al. 1998). A05 calculated the thermal, chemical, and ionization balance for an $H^+$ region in gas pressure equilibrium with a PDR ($P_{th}$=constant), but did not include a magnetic field in the equation of state.

### *2.2.4* Combining $H^+$ and PDR Emission Line Diagnostics

The influence of the magnetic field can be constrained by considering the ratio of an $H^+$ region to PDR emission line. The intensity of an $H^+$ region emission-line will depend on $U$, $n(H^+)$, and $T_*$ (for a constant abundance). However, the intensity of a PDR emission-line depends on $n_{PDR}$ and $T_{PDR}$, which are determined by the magnetic fields effects on the equation of state. Therefore, the ratio of an $H^+$ emission-line to a PDR emission line will scale with the equation of state.

Ideally, the best diagnostic ratio would be one where both the $H^+$ and PDR emission-line emerged from the same element, since such a ratio would not depend on the abundance ratio of two elements. One such ratio that can be measured in the infrared is [O III] (88.3 or 51.8 μm) to [O I] (63.2 or 145.5 μm). A given $n(H^+)$, $T_*$, and $U$, determines the [O III] emission, while the [O I] emission-lines form largely in the PDR. Therefore, the quantities $n_{PDR}$ and $T_{PDR}$ and the resulting [O III]/[O I] ratio depends only on the equation of state.

# 3 Calculation Details

We use version 05.07.06 of the spectral synthesis code Cloudy (Ferland et al. 1998) to perform our calculations. The treatment of PDR physics is discussed in A05.

We varied four parameters in our calculations. These were $U$, $T^*$, $n(H^+)$, and the equation of state. The range of parameters were (with increments in parentheses): $U$ =0.03-0.0003 (1 dex), $T^*$ = 30,000–50,000 K (5,000 K), and $n(H^+)$ = 30-3,000 $cm^{-3}$ (1 dex). In all our calculations, the $H^+$ region and PDR were connected by assuming either constant density (which would be the case if the magnetic field or turbulence dominates the pressure) or constant thermal pressure (no magnetic field). Our results therefore represent the two possible limiting cases. For simplicity, we assumed a plane parallel geometry.

We use gas and grain abundances representative of the Orion environment, a typical $H^+$ region on the surface of a molecular cloud. The complete set of abundance used is given in Baldwin et al. (1996). A few by number are: He/H = 0.095, C/H= $3\times10^{-4}$; O/H= $4\times10^{-4}$, N/H= $7\times10^{-5}$, Ne/H= $6\times10^{-5}$, and Ar/H= $3\times10^{-6}$. We have assumed S/H= $2\times10^{-6}$ based on observations of starburst galaxies by Verma et al. (2003). Grains in the Orion environment are known to have a larger then ISM size distribution (Cardelli et al. 1989), as are those in the starburst galaxies studied by Calzetti et al. (2000). We therefore use a truncated MRN size distribution (Mathis et al. 1977), weighted toward larger grains (Baldwin et al. 1991). PAHs are also included in our calculation with a size distribution given in Bakes & Tielens (1994). PAHs are known to primarily exist in regions of atomic hydrogen (see, for instance, Giard et al. 1994). We assume that the number of carbon atoms in PAHs per hydrogen, $n_C(PAH)/n_H$ is $3\times10^{-6}$. The PAH abundance is then scaled by the ratio $H^0/H_{tot}$, to produce low values in ionized and molecular regions (A05).

We consider two sources of ionization, a stellar continuum and cosmic rays. For the stellar continuum, we used the WMBasic O star atmosphere models of Pauldrach, Hoffman, & Lennon (2001). We use the tabulated supergiant continuum with solar metallicity. These continua were also used by Morisset (2004) in his determination of $T_*$ and $U$ in our galaxy. We treat cosmic ray processes as described in A05. We include primary ionizations, with an ionization rate $\xi = 5\times10^{-17}$ s$^{-1}$, and secondary ionizations caused by energetic electrons ejected by cosmic rays. Suchkov et al. (1993) find that the cosmic-ray ionization rate in the starburst galaxy M82 is enhanced over galactic by a factor of 200. Tests show that this level of enhancement has little effect on the [O I] emission from the PDR.

With this set of parameters, we determine the ionization and thermal balance and the resulting spectrum. Our calculation begins at the illuminated face of the $H^+$ region, continues through the PDR, and ends deep in the molecular cloud, at a visual extinction ($A_V$) of 100.

# 4 Results

Figures 1-4 show the results of our calculations. The [Ne III] 15.6μm to [Ne II] 12.8μm line intensity ratio is the x-axis and the y-axis is the ratio of [O III] 88.3μm to [O I] 63.2μm. As mentioned above, the first ratio is a $T_*$ and $U$ indicator, and the second is primarily sensitive to $n_{PDR}$ and $T_{PDR}$.

Figure 1 shows all of our calculations (constant gas pressure, where the magnetic field is small, and constant density, appropriate if the magnetic field dominates) on a single diagram. Additionally, we placed a sample of observations taken from the literature, primarily being Giveon et al. (2004), Malhotra et al. (2001), Verma et al. (2003), and Morrisset (2004). Figure 1 shows that constant density generally results in a larger [O III]/[O I] ratio because the PDR density is lower in this case, and lower density produces less [O I] emission. Unfortunately, there is clear overlap between the constant gas pressure and constant density calculations. Additionally, many of the observations shown here fall in this overlap region. For [Ne III]/[Ne II] vs. [O III]/[O I] to be a useful diagnostic, there would have to be a clear gap between the constant pressure and constant density results. Therefore, we conclude that observations of these four emission lines are not enough to determine unambiguously the equation of state.

Figures 2-4 again shows the same diagnostic ratios, but this time the dependence of [Ne III]/[Ne II] vs. [O III]/[O I] is shown for a given $U$ and $n(H^+)$. The dependence of [O I] emission on $n_{PDR}$ and $T_{PDR}$ is clearly seen in Figures 2-4.

The [O I] emission is weaker relative to [O III] in the constant density, high magnetic field case. This is due to the temperature in the PDR. The constant density case will have a lower PDR density than the constant pressure case. A lower density PDR will typically also have a lower temperature (Tielens & Hollenbach 1985, Figure 10a). A lower temperature means fewer collisional excitations of ground state oxygen. This makes the [O III]/[O I] ratio sensitive to the equation of state. For almost all combinations of $n(H^+)$, $T_*$, and $U$, the difference in the [O III]/[O I] ratio for the two equations of state is greater than 1 dex. The separation is smaller in low-$U$, low $n_H$, clouds because a significant fraction of the [O I] emission comes from the $H^+$ region.

Overall, our results show that the equation of state can be determined, but only if $U$ and $n_H$ are known. This is possible, using the diagnostics mentioned in Section 2. However, such an analysis would only be strictly valid for a single $H^+$ region adjacent to a PDR, where a single set of parameters exists. In the case of extragalactic observations, ensembles of $H^+$ regions and PDRs, each with their own values of $U$ and $n(H^+)$, will be seen in a single observation. In this case, the entire galaxy can be parameterized by an effective $U$ and $n_H$ but the physical meaning of these is unclear. We therefore find that our analysis cannot be used to determine the equation of state (or the role of the magnetic field) for extragalactic star-forming regions.

# 5 Applications

Our results have several applications. The most obvious application is to local star-forming regions where spectral data from a single $H^+$ region and PDR can be observed. If overall pressure equilibrium is assumed, then a theoretical calculation can deduce $U$, $T_*$, and $n(H^+)$ from the observed spectrum. The magnetic field is then a free parameter, which can be varied until the observed [O III]/[O I] ratio is reproduced. Both SOFIA and Herschel will be capable of making high spectral and spatial resolution observations of galactic star forming regions, many without magnetic field measurements. Assuming that shocks are unimportant (which could also be determinable from the spectrum), then our methods provide a way to study the effects of magnetic fields in star-forming regions throughout the galaxy. In regions where magnetic field observations do exist, such as S106 or NGC 6334, then we can test the validity of our technique.

Future observations could interpret measurements of the [O III]/[O I] and [Ne III]/[Ne II] ratio for a large sample of galaxies in terms of the equation of state. Even though there is an overlap region in Figure 1 where either isobaric or isochoric models are possible, there are also regions where the two equation of state are clearly separated. If these two ratios were measured for a wide range of galaxies, then we may find cases where the observations are best explained by a particular equation of state. Again, SOFIA and Herschel can lead the way in determining what equation of state best reproduces observation.

## 5.1 The Equation of State of Orion

As a test of the methods outlined in this work, we applied our calculations to the line of sight directly behind the Trapezium Cluster of Orion (Figure 5). Orion has been extensively studied in the infrared (see for instance Tauber et al 1994). Additionally, both the stellar continuum and distance of the Trapezium to the man $H^+$ ionization front are well determined (O'Dell 2001 and references therein). We know that layer of gas between the Trapezium and Earth, Orion's Veil, has a magnetic field strength of ~100μG (Abel et al. 2004). Because the Veil is associated with the Orion complex, we would expect the magnetic field to play some role in the equation of state connecting M42 to the parent molecular cloud OMC-1. Faraday rotation measurements of Rao et al. (1998) place upper limits to the magnetic field in the $H^+$ region of < 350μG.

The calculation details are essentially the same as those given in Section 3, with a few exceptions. Essentially, our calculations for the $H^+$ region follow those presented in Ferland (2001), which is based on the observations and theoretical calculations given in Baldwin et al. (1991). Our model is a plane-parallel slab illuminated on one side by the modified Kurucz LTE atmosphere described by Rubin et al. (1991). The flux of hydrogen ionizing photons $\phi(H^0)$ = $10^{13}$ photons $cm^{-2}$ $s^{-1}$, and $n(H^+)= 10^{3.8}$ $cm^{-3}$. This combination of $\phi(H^0)$ and $n(H^+)$ corresponds to $\log[U] = -1.6$. Our calculations extend to $A_V$ = 10 mag, which sufficiently accounts for all [O I] emission in the PDR. We calculated the [Ne III]/[Ne II] ratio for this set of conditions and found [Ne III]/[Ne II] = 1.10, in relatively good agreement with the observed value of 0.86 (Simpson et al. 1998; see Table 1).

Since we know the properties of the $H^+$ region and the observed spectrum of the PDR we can now determine the equation of state. We specify a magnetic field in the $H^+$ region, and assume constant pressure with a magnetic field – gas density scaling law with $\kappa = 2/3$ ($B \sim n^{2/3}$). We then will find what magnetic field reproduces the observed [O III]/ [O I] intensity ratio. For the [O III] 88 μm and [O I] 63 μm intensity, we use position 1 of Furniss et al. (1983), which gives an intensity of 0.11 ± 0.08 erg $cm^{-2}$ $s^{-1}$ for [O III] and 0.46 ± 0.07 erg $cm^{-2}$ $s^{-1}$ for [O I]. This yields a ratio of 0.06 – 0.48, with a mean value of 0.23.

Figure 6 and Table 1 shows our results. For low values of the magnetic field ($10^{-7} < B(H^+) < 10^{-4.5}$ G) gas pressure dominates in the $H^+$ region and PDR. The calculations for this $B(H^+)$ range are therefore identical, with an [O III]/[O I] ratio of ~0.1. For $B(H^+) > 10^{-4.5}$ G, the magnetic pressure in the PDR becomes important. The magnetic pressure reduces $n_{PDR}$, which reduces the average temperature in the $O^0$ region (Figure 7). As mentioned in Section 4, the lower temperature leads to less excitation of the $^3P_1$ level of $O^0$ and therefore less emission. This increases the [O III]/[O I] ratio. The magnetic field strength that best reproduces the [O III]/[O I] ratio is $B(H^+) = 10^{-3.95}$ G. For this magnetic field, $n_{PDR} = 1.5\times10^5$ $cm^{-3}$. Equation 5 combined with Figure 7 ($T_{PDR}$ = 220 K) and $T(H^+)$ (9,000 K from our models) yields a $\beta$ of ~1.2, meaning that magnetic pressure and thermal pressure in the PDR are roughly equal.

Although our results are much closer to constant gas pressure than constant density, we find that the effects of the magnetic field on the equation of state cannot be neglected. Figure 6 shows that the observed [O III]/[O I] ratio has a value that occurs in the transition region between the limits where thermal and magnetic pressure dominate. Orion is often used as a test case for understanding physical processes in extragalactic environments. Our calculations show that the magnetic field is important to the equation of state in Orion, meaning it is likely important (and therefore not negligible) in extragalactic star-forming regions.

The derived physical properties of the $H^+$ region and PDR are consistent with previous studies. Our derived PDR density of ~$10^5$ $cm^{-3}$ is consistent with Tielens & Hollenbach (1985). Our derived $B(H^+)$ falls well below the Rao et al. (1998) limit. If the magnetic ($B^2/8\pi$) and turbulent ($1/2\rho v^2$) energy densities are in equipartition, then the turbulent broadening in the $H^+$ region is ~2.1 km $s^{-1}$, well below the amount of unexplained broadening observed in the $H^+$ region emission lines of M 42 (Castañeda 1988, O'Dell 2000). If we assume magnetic – turbulent equipartition and a density-magnetic field scaling law of $\kappa = 2/3$, the turbulent broadening will be proportional to $n^{1/6}$. For our derived PDR density of $1.5\times10^5$ $cm^{-3}$, this yields a turbulent broadening in the PDR of ~3.3 km $s^{-1}$. This is an appreciable fraction of the observed [C II] 158 μm and [O I] 63 μm linewidths of 5.4 km $s^{-1}$ and 6.8 km $s^{-1}$, respectively (Boreiko & Betz 1996).

# 6 Conclusions

1. We have investigated whether ratios of emission-lines from the $H^+$ region and PDR can determine the equation of state in star-forming regions. We found that the ratio of an $H^+$ region emission line to a PDR emission line, when plotted against a line ratio that is a diagnostic indicator of the intensity of the radiation field in the $H^+$ region, is sensitive to the equation of state. The most promising of these is the ratio of [O III]/[O I], since it is independent of abundance.
2. We find that the methods outlined here will are only strictly valid in the limited case of single $H^+$ regions adjacent to a PDR where a single set of physical conditions apply. In this case, knowledge of $U$, $T_*$, and $n(H^+)$, combined with our diagnostic diagrams can determine the equation of state and therefore estimate $\beta$. However, for galaxies, which consist of ensembles of $H^+$ regions and PDRs, our results are ambiguous since a region of overlap exists where constant density or pressure models can explain the observations.

Therefore, the diagnostic line ratios outlined here cannot determine the equation of state or the role of magnetic fields in extragalactic star-forming regions using current observations. If future observations identify galaxies that are not in the overlap region, then our methods could determine the equation of state.

3. We applied our calculations to the Orion Complex, along a ray starting at the Trapezium and going through both M 42 and the PDR. We derive a magnetic field in the $H^+$ region of $10^{-3.95}$ μG. We find that the magnetic and thermal pressure in the PDR are roughly equal, with $\beta \sim 1.2$. These parameters are in good agreement with other estimates. We also find that, if magnetic and turbulent energies are in equipartition, then the amount of line broadening due to the magnetic field is insignificant in the $H^+$ region, but may be important in the Orion PDR. Note that neither the constant density or constant gas pressure cases would produce the deduced PDR density (Tielens & Hollenbach, 1985) given the measured $H^+$ region density –This is a true statement.

Acknowledgements: NPA would like to acknowledge the Center for Computational Sciences at the University of Kentucky for computer time and financial support through a CCS fellowship. We thank NSF, NASA, STScI, and Spitzer for support through grants AST 0307720, NNG05GG04G, HST-AR-10636.01, HST-AR-10316, HST-AR-10652, and HST-AR-10653.

# 8 Figures

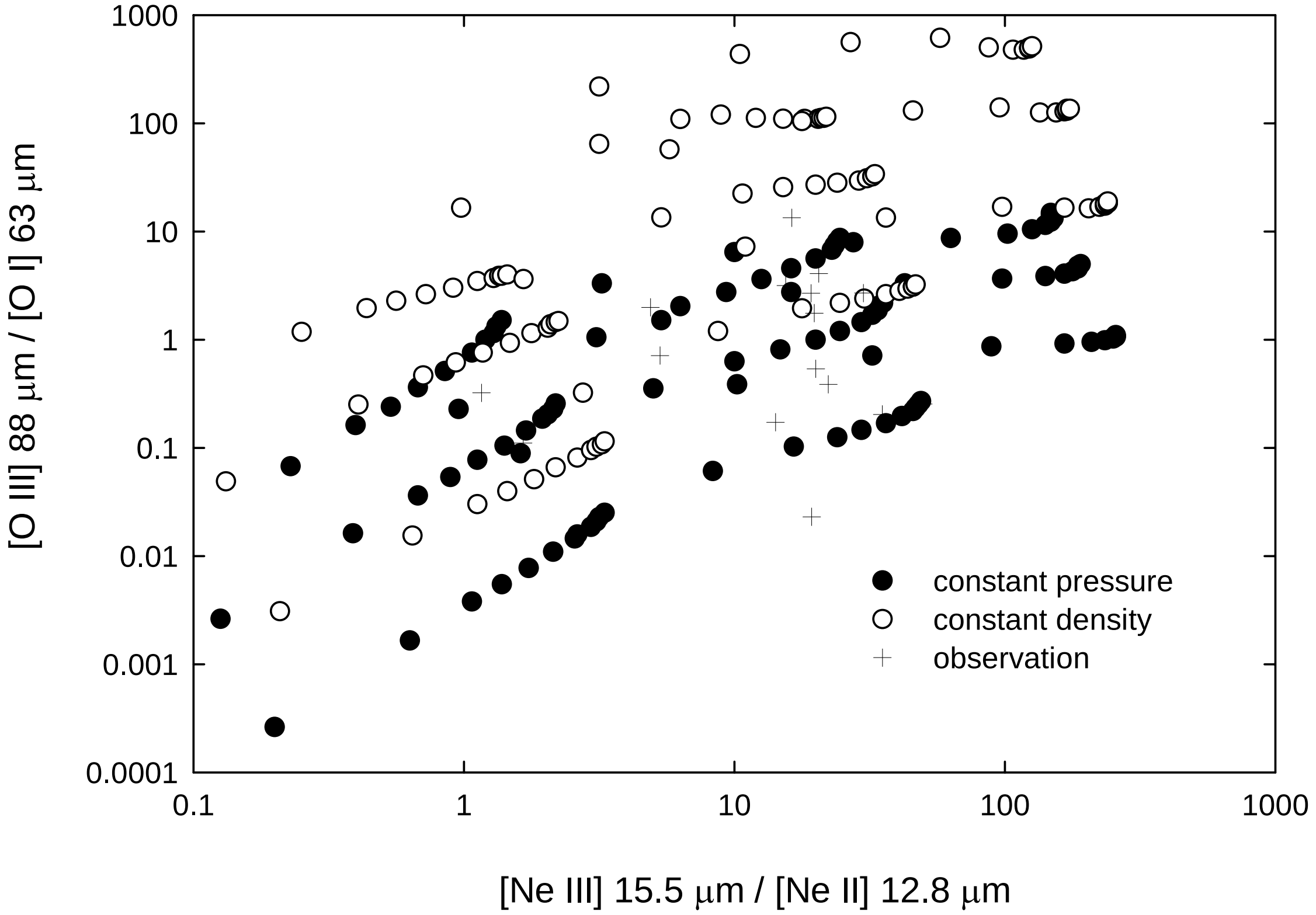

Figure 1 Ratio of [Ne III]/[Ne II] vs. [O III]/[O I] for $P^{mag} >> P^{th}$ (open circles), and $P^{mag} = 0$ (filled circles) for all combinations of $U$ and $n(H^+)$ considered. For most of the parameter space, there is a region of overlap where the equation of state could be dominated either by gas or magnetic pressure. The lower and upper regions are only reproduced by constant pressure or constant density, not both. If extragalactic observations of [O III]/[O I] in star-forming regions fall in these extremes, the deriving the equation of state is possible. Otherwise, the results are ambiguous.

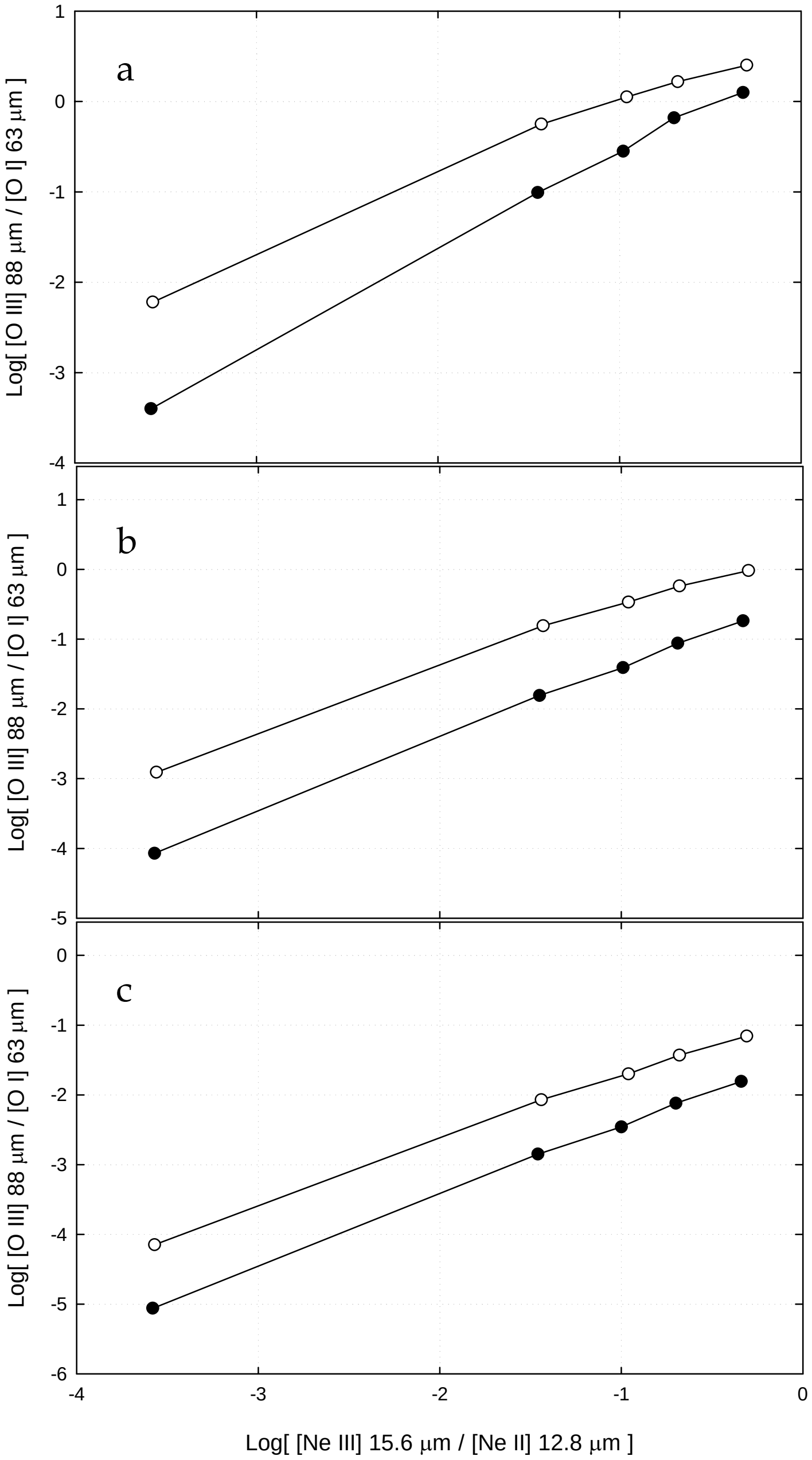

Figure 2 Ratio of [Ne III]/[Ne II] vs. [O III]/[O I] for $P^{mag} >> P^{th}$ (open circles), $P^{mag} = 0$ (filled circles), $U = 10^{-3.5}$, and $n(H^+) = 10^{1.5}, 10^{2.5}, 10^{3.5}$ cm$^{-3}$ (a, b, and c, respectively). For a given $U$ and $n(H^+)$, the [O III]/[O I] ratio is sensitive to $\beta$. Our calculations, therefore, allow a way to determine $\beta$ in constant pressure environments.

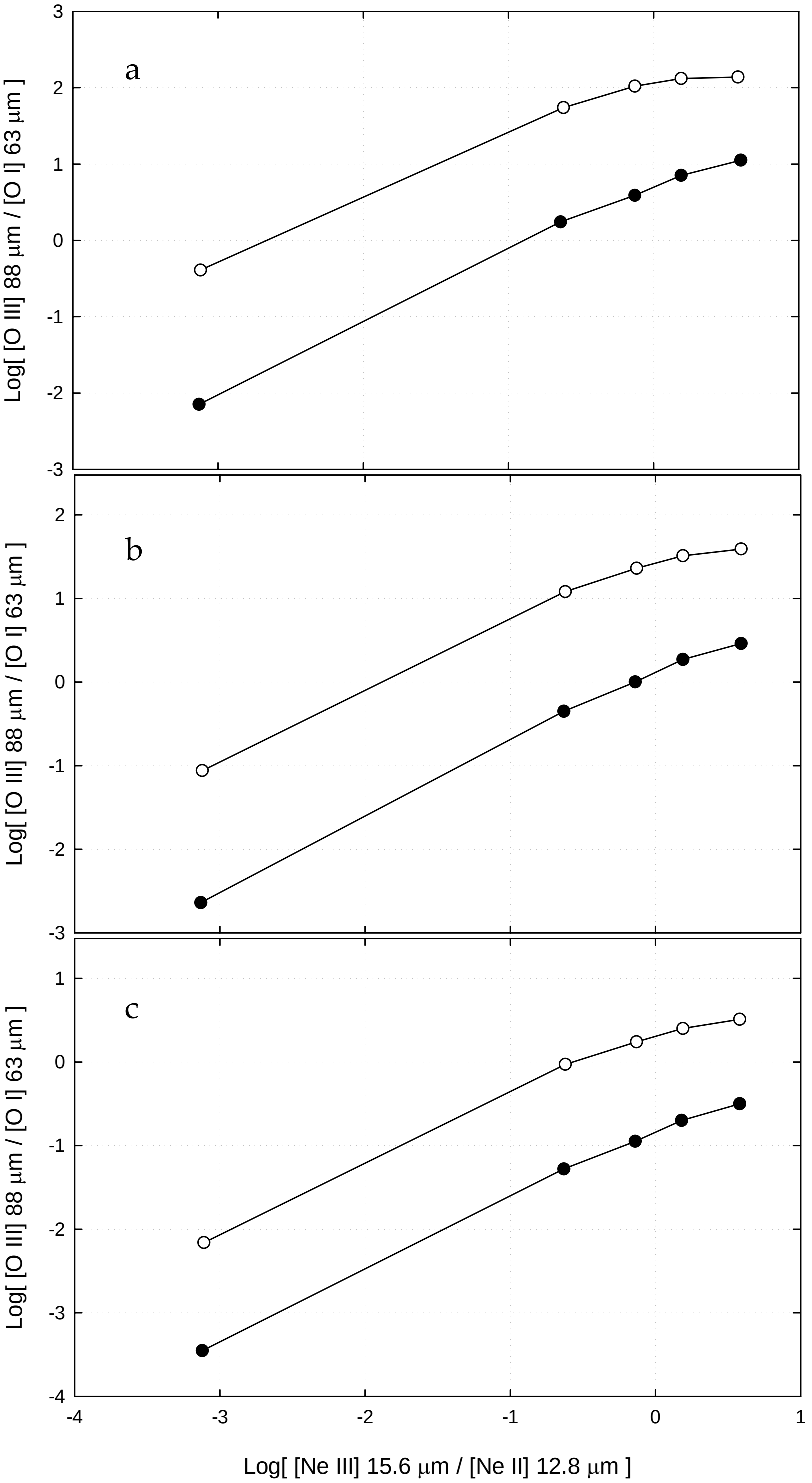

Figure 3 Ratio of [Ne III]/[Ne II] vs. [O III]/[O I] for $P^{mag} >> P^{th}$ (open circles), $P^{mag} = 0$ (filled circles), $U = 10^{-2.5}$, and $n(H^+) = 10^{1.5}, 10^{2.5}, 10^{3.5}$ $cm^{-3}$ (a, b, and c, respectively). For a given $U$ and $n(H^+)$, the [O III]/[O I] ratio is sensitive to $\beta$. Our calculations, therefore, allow a way to determine $\beta$ in constant pressure environments.

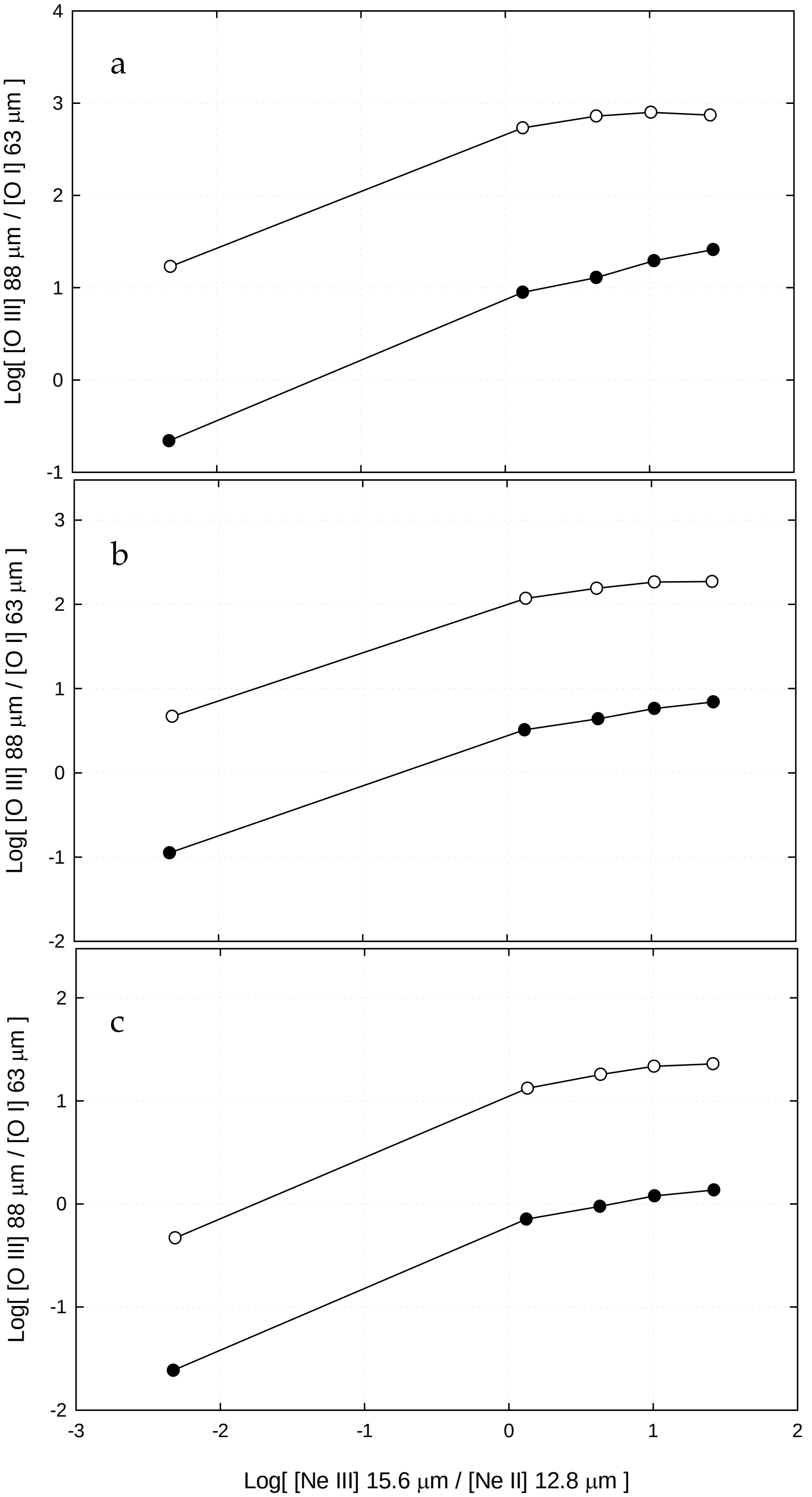

Figure 4 Ratio of [Ne III]/[Ne II] vs. [O III]/[O I] for $P^{mag} >> P^{th}$ (open circles), $P^{mag} = 0$ (filled circles), $U = 10^{-1.5}$, and $n(H^+) = 10^{1.5}, 10^{2.5}, 10^{3.5}$ cm$^{-3}$ (a, b, and c, respectively). For a given $U$ and $n(H^+)$, the [O III]/[O I] ratio is sensitive to $\beta$. Our calculations, therefore, allow a way to determine $\beta$ in constant pressure environments.

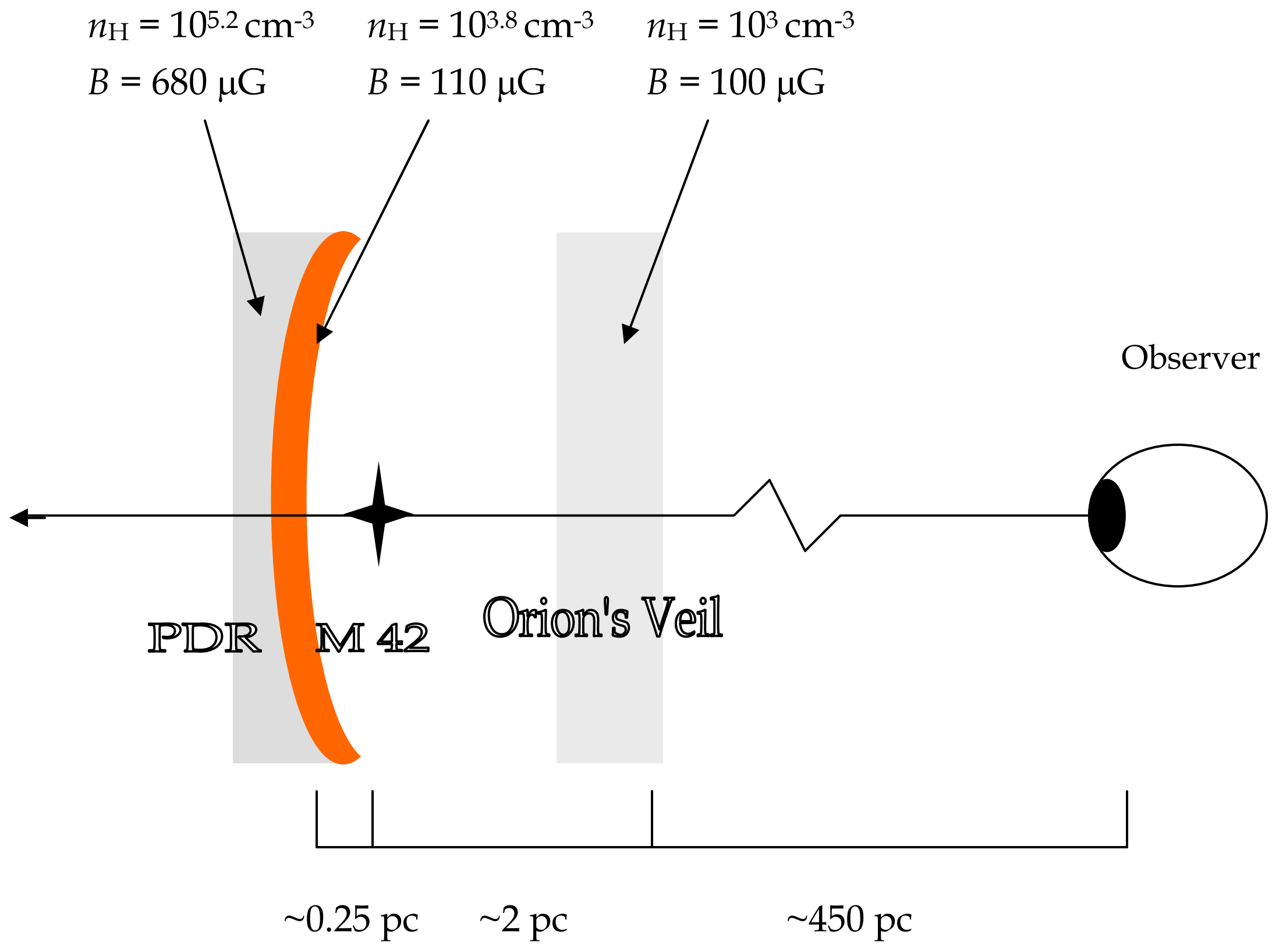

Figure 5 Geometry of the Orion Environment along the line of sight to the Trapezium Cluster. Orion's Veil is the absorbing screen in front of the Trapezium Abel et al. 2004, 2006). Beyond the Trapezium Cluster is the Orion Nebula (M 42), and beyond the hydrogen ionization front is the Orion PDR / molecular cloud.

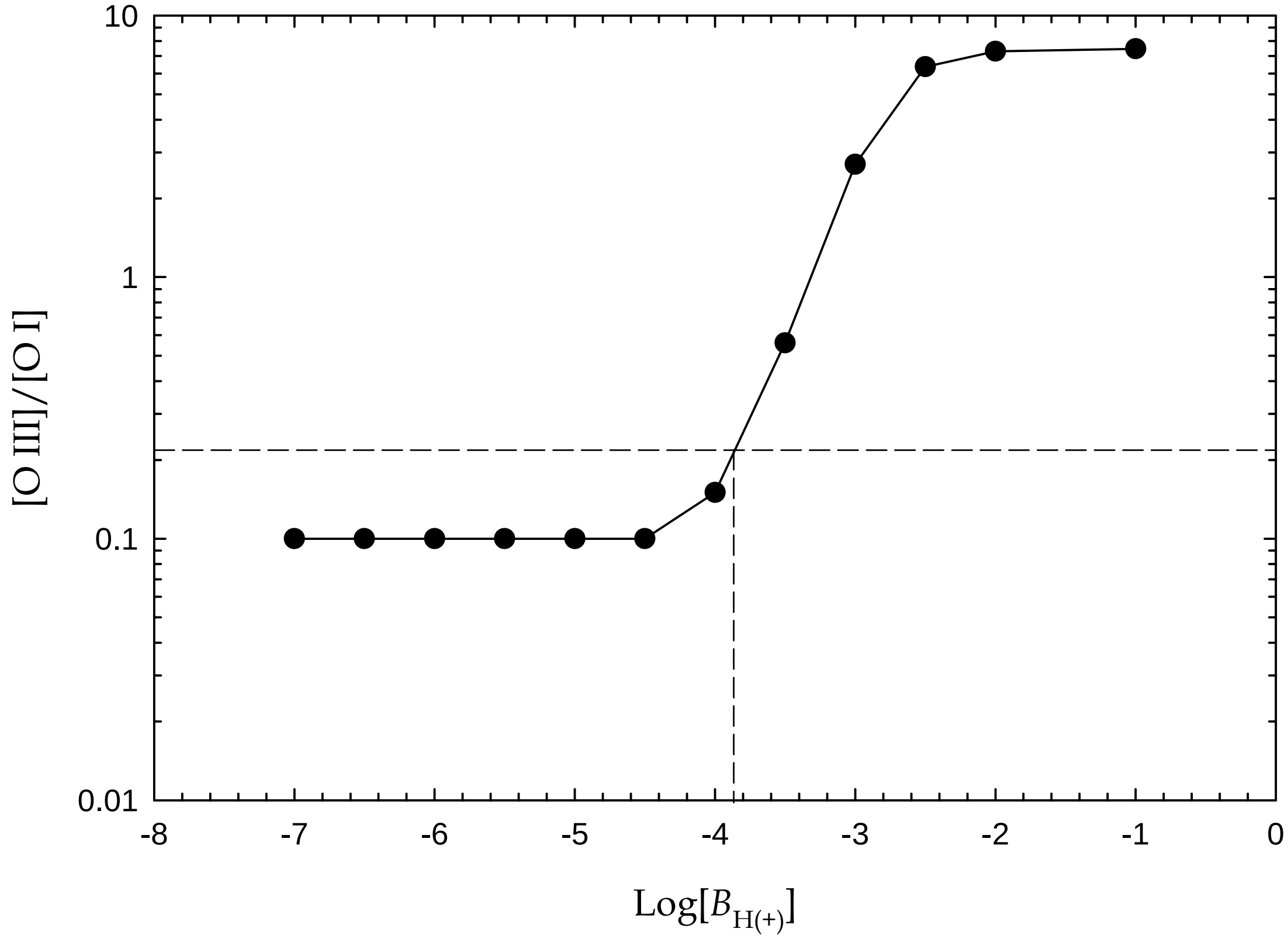

Figure 6 The dependence of the [O III]/[O I] ratio on the magnetic field in the Orion $H^+$ region, with a magnetic field scaling law $\kappa$ of 2/3. The horizontal dashed line is the observed [O III]/[O I] ratio from Furniss et al (1983). For low $B(H^+)$ ($<10^{-4}$ G), the magnetic pressure is much less than the gas pressure and therefore plays no role in the equation of state. For $B(H^+) > 10^{-4}$ G, the magnetic pressure starts to dominate over gas pressure in the PDR. This lowers the PDR density and [O I] emission. The vertical dashed line indicates the value of $B(H^+)$ which reproduces the observed [O III]/[O I] ratio.

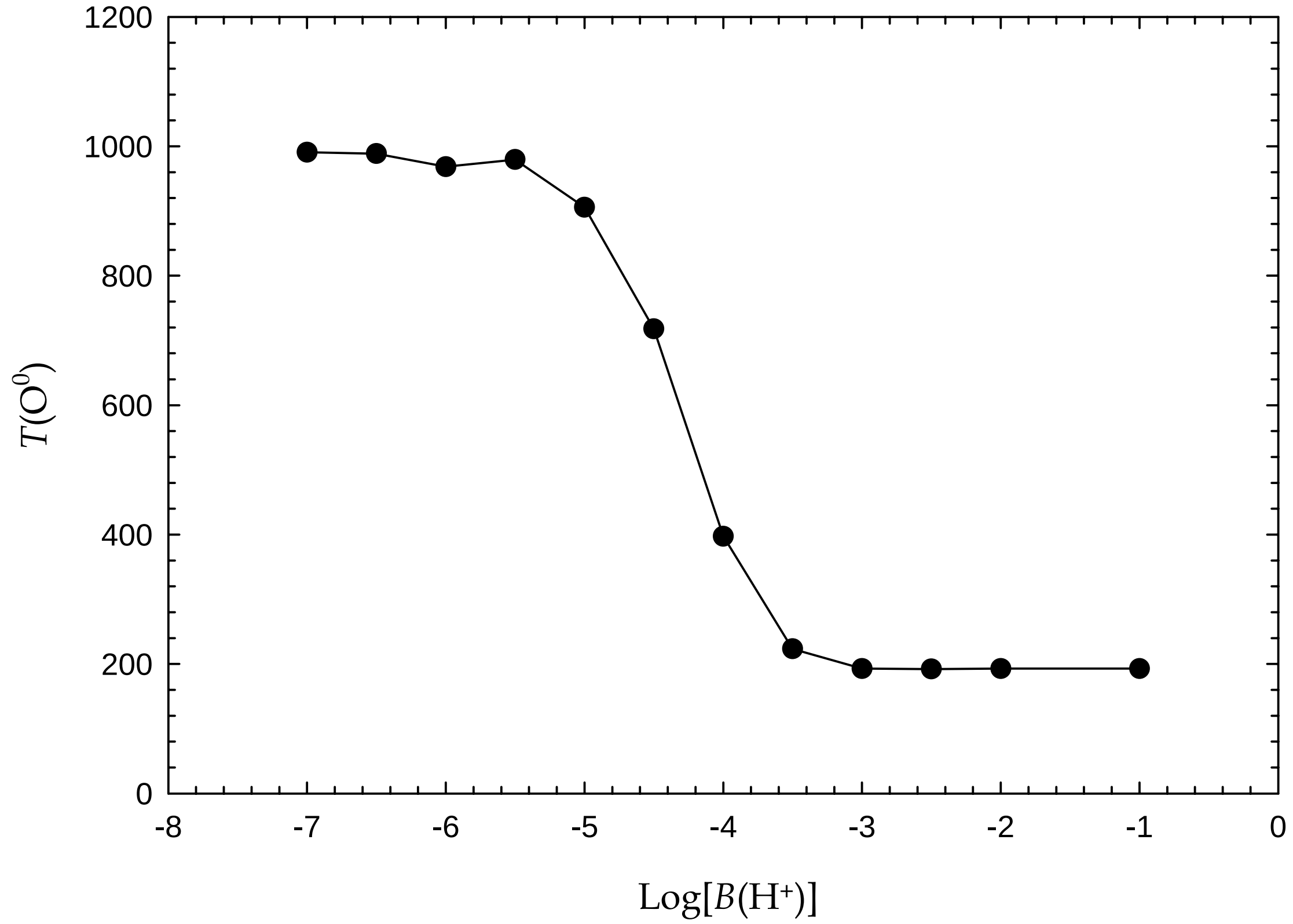

Figure 7 The temperature of the $O^0$ region on the magnetic field in the Orion $H^+$ region, with a magnetic field scaling law $\kappa$ of 2/3. For low $B(H^+)$ (<$10^{-4}$ G), the magnetic pressure is negligible, making the PDR density and temperature nearly constant. For $B(H^+) > 10^{-4}$ G, the magnetic pressure starts to dominate over gas pressure in the PDR. This decreases the density in the PDR and therefore the temperature in the $O^0$ region. This decrease in temperature also decreases [O I] emission, which explains the increase in [O III]/[O I] shown in Figure 6.

Table 1 Comparison of Orion $H^+$/PDR Model with Observation

| Line Ratio | $I_{obs}$ (erg cm$^{-2}$ s$^{-1}$) | $I_{calc}$ (erg cm$^{-2}$ s$^{-1}$) | Reference |
|---|---|---|---|
| $\frac{\text{[Ne III] } 15.6\mu\text{m}}{\text{[Ne II] } 12.8\mu\text{m}}$ | 0.86 | 1.10 | Simpson et al. (1998) |
| $\frac{\text{[S IV] } 10.5\mu\text{m}}{\text{[S III] } 18.7\mu\text{m}}$ | 0.30 | 0.57 | Simpson et al. (1998) |
| $\frac{\text{[Ar III] } 9.00\mu\text{m}}{\text{[Ar II] } 6.98\mu\text{m}}$ | 3.90 | 2.86 | Simpson et al. (1998) |
| $\frac{\text{[O III] } 88.3\mu\text{m}}{\text{[O III] } 51.8\mu\text{m}}$ | 0.15 | 0.14 | Furniss et al. (1983) |
| $\frac{\text{[O III] } 88.3\mu\text{m}}{\text{[O I] } 63.2\mu\text{m}}$ | 0.23 | 0.19 | Furniss et al. (1983) |